\documentclass{INTERSPEECH2023}

\title{PoCaPNet: A Novel Approach for Surgical Phase Recognition Using Speech and X-Ray Images}
\name{Kubilay Can Demir$^1$, Tobias Weise$^{1,2}$, Matthias May$^3$, Axel Schmid$^3$, Andreas Maier$^2$, Seung Hee Yang$^1$}

\address{
  $^1$SLU Lab. Friedrich-Alexander-Universität Erlangen-Nürnberg, Germany\\
  $^2$Pattern Recognition Lab. Friedrich-Alexander-Universität Erlangen-Nürnberg, Germany \\
  $^3$Department of Radiology Friedrich-Alexander-Universität Erlangen-Nürnberg, Germany}
\email{kubilay.c.demir@fau.de, seung.hee.yang@fau.de}

\begin{document}

\maketitle
 
\begin{abstract}
Surgical phase recognition is a challenging and necessary task for the development of context-aware intelligent systems that can support medical personnel for better patient care and effective operating room management. In this paper, we present a surgical phase recognition framework that employs a Multi-Stage Temporal Convolution Network using speech and X-Ray images for the first time. We evaluate our proposed approach using our dataset that comprises 31 port-catheter placement operations and report $82.56$ \% frame-wise accuracy with eight surgical phases. Additionally, we investigate the design choices in the temporal model and solutions for the class-imbalance problem. Our experiments demonstrate that speech and X-Ray data can be effectively utilized for surgical phase recognition, providing a foundation for the development of speech assistants in operating rooms of the future. 

\end{abstract}

\section{Introduction}

Modern operating rooms (ORs) are optimized for better patient care and the most effective utilization of medical resources. Advances in technology are presented in ORs with cutting-edge surgical tools, monitoring, and navigation systems. These systems enable physicians to perform more complex surgical procedures with a high success rate that was not possible before~\cite{maier2017surgical}. Simultaneous to these advances, the amount of data created by modern medical systems is increasing~\cite{vercauteren2019cai4cai}. This data is necessary for the successful execution of the operation and needs to be processed by the medical personnel after or during operations. It has been proposed that intelligent systems processing this growing amount of information and projecting it in the correct time and format will be vital in the future of ORs~\cite{cleary2004or2020, rattner2003advanced}.

Surgical phase recognition (SPR) is a topic of automatically extracting semantic information from ongoing or recorded surgical operations by recognizing different predefined phases~\cite{lalys2014surgical}. The highest level actions performed in the operating room, such as anesthesia, sterilizing, or cutting, are referred to as surgical phases. Robust estimation of these phases is a prerequisite for the development of the envisioned context-aware intelligent assistants in the OR.  

The majority of studies in SPR focused on \textit{laparoscopic cholecystectomy}, removal of the gallbladder, through endoscopic videos \cite{twinanda2016endonet, czempiel2020tecno, jin2020multi, gao2021trans, nwoye2022rendezvous}. Microscopic videos are another popular source of information used mainly for cataract surgeries~\cite{primus2018frame, xia2021against, qi2019deep}. Sensory data from robotic surgeries are considered together with surgical videos in several studies~\cite{paysan2021self, nwoye2022rendezvous}. As these modalities are not used in every operation, it is not possible to cover all types of operations with endoscopic and microscopic videos. Moreover, these data are typically recorded inside or near the body and do not contain information about the environment of the OR. Despite the growing interest in SPR, the use of speech and audio data has received little attention~\cite{Demir2022}. \textit{Guzm{\'a}n-Garc{\'\i}a et al.}~\cite{guzman2021speech} extracted Spanish transcriptions from 15 online education videos on \textit{laparoscopic cholecystectomy} and achieved $82.95$ \% accuracy in surgical phase recognition. \textit{Seibold et al.}~\cite{seibold2022conditional} used discrete segments from the German audio dataset of five \textit{total hip arthroplasty} operation for the phase recognition task and recognized seven phases with $95.60$ \% accuracy. We hypothesize that using speech and audio is a necessary direction in SPR as it can open the way for various applications based on natural language processing (NLP) and enable the development of interactive smart assistants in ORs. 

In this study, we propose a novel approach for SPR using speech and X-Ray images collected during port-catheter placement operations. Our method utilizes Multi-Stage Temporal Convolutional Network (MS-TCN)~\cite{farha2019ms} architecture and past estimations for temporal modeling, leverages wav2vec 2.0 XLSR-53~\cite{conneau2020unsupervised} representations for speech signals and TorchXRayVision~\cite{Cohen2022xrv} representations for X-Ray images.

The contribution of our work as follows:
\begin{itemize}
\item To best of our knowledge, it is the first approach in SPR that utilizes speech data from entire surgical operations and combination with X-Ray images.
\item We analysed usage of positional encodings and previous estimates for integrating long term temporal information.
\item We analysed class-imbalance problem with re-weighting and modified loss functions for recognizing short duration surgical phases.
\end{itemize}

Our study is organized as follows: we explain our proposed framework in Section~\ref{sec:method}; we report our findings in Section~\ref{sec:experiments}, and we give our conclusion in Section~\ref{sec:conc}.

\section{Proposed Method}
\label{sec:method}

\begin{figure*}[t]
  \centering
  \includegraphics[width=\textwidth]{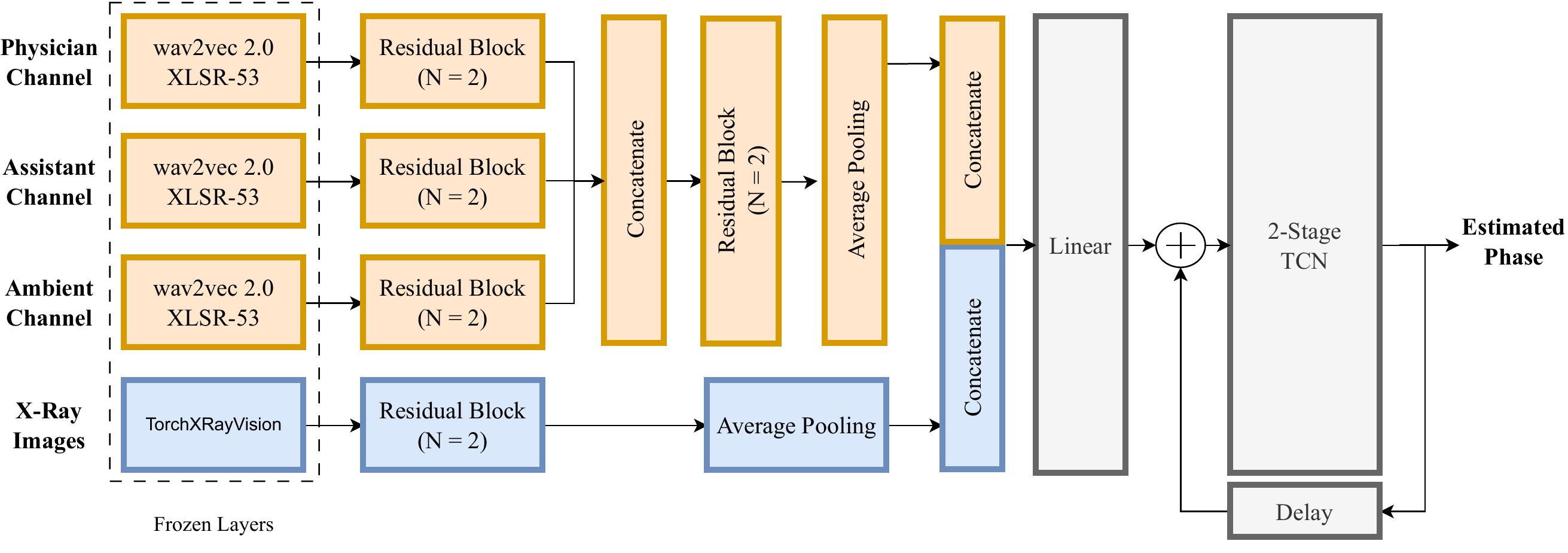}
  \caption{The framework of the proposed model. The orange-colored blocks represent the audio branch, blue-colored blocks represent the visual branch and grey-colored blocks represent the temporal model. \textit{N} shows the number of residual blocks.}
  \label{fig:model}
\end{figure*}

\subsection{Intervention}

Port-catheter placement is a frequently applied minimally-invasive procedure in the radiology department. The main purpose of this operation is to place a port under the skin of the chest, which is connected via a catheter to the large veins emptying into the heart. This operation prevents injuries to small vessels after repetitive infusions during treatments such as chemotherapy~\cite{gonda2011principles}. The procedure is typically performed by a single physician and an assistant. The duration of an operation varies between half an hour to three hours depending on the experience level of the medical personnel, complications during the operation or the general condition of the patient.

\subsection{Feature Extraction Backbone}
For speech and audio signals, self-supervised approaches such as wav2vec 2.0~\cite{baevski2020wav2vec} have been widely used. To create better representations for languages with little available data, the wav2vec 2.0 model extended to multi-language setting in XLSR~\cite{conneau2020unsupervised}. A large variant XLSR-53 is trained with 50k hours of public data from 53 languages, including German which is the language of the experiment corpus. Therefore, we chose XLSR-53 as the backbone model for our framework. In our study, we used the output of the final Transformer layer. The motivation for this choice is to include maximum temporal information in the feature vectors. In the windowing step, we used seven-seconds casual Hann window with a one-second hop length. The seven-second long audio data of each channel is used as the input to the feature extraction backbone. 

For X-Ray images, we used Densenet121~\cite{huang2017densely} model pre-trained on all publicly available chest X-Ray datasets. The details of the pre-trained model and the corresponding TorchXRayVision library are released in~\cite{Cohen2022xrv}. The pretraining setting and datasets are closely relevant to the port-catheter placement intervention as chest X-Rays are used. To have the same temporal resolution as the speech and audio features, we extracted features from seven X-Ray images, which correspond to seven seconds at a $1 fps$ rate, and shifted a single image at each time step. Both wav2vec 2.0 XLSR-53 and Densenet121 models held frozen during training. 

\subsection{Temporal Model}
Modeling temporal relations is a vital component of SPR architectures. To have a large receptive field, we used a two-stage TCN network in our temporal model. Although Transformer networks became very popular and achieved state-of-the-art results in many applications, we chose TCN architecture in our network due to limited available training data. We used past estimates in an auto-regressive manner to further increase the receptive field of the network. The effect of our design choices are experimented in Section~\ref{sec:exp_temp}.

Our model runs at a rate of one second, i.e. an estimation is made every second utilizing features from a window of seven seconds as explained in the previous section. The overview of our whole proposed framework is depicted in Figure~\ref{fig:model}. Our source code is available at: \url{https://github.com/kubicndmr/PoCaPNet}.

\section{Experiments}
\label{sec:experiments}

\subsection{Dataset}
In our study, we used the PoCaP Corpus containing 31 port-catheter placement operations recorded in the Radiology Department of University Hospital Erlangen, Germany~\cite{demir2022pocap}. In this dataset, the physician and the medical assistant wore Sennheiser XSW 2 ME3-E wireless headsets for the audio recording. All conversations in the dataset are held in German. Additionally, the internal microphone of a single GoPro Hero 8 camera, which is initially set up for easing annotation work, is also included in the dataset as the ambient microphone channel. All channels are aligned and then re-sampled at $16kHz$. X-Ray images are captured from the output of the X-Ray machine at a $1 fps$ rate. The data set is unfortunately not publicly available due to local laws for protecting the data privacy of the patients and the medical personnel. To the best of our knowledge, any other multi-modal data set of any operation type containing full-length speech signals is not existing for repeating our experiments. Thus, we can only provide our results with our in-house dataset.  

The recordings suffer significant data loss during the six operations. Data loss happened during data recording for a variety of causes, such as software failure, a change in operating personnel, or a recording error. Thus, we excluded these recordings and used the remaining 25 operations for our experiments. For the training, validation, and test set separation, we employed the conventional $60-20-20$ percent random split. 

The dataset contains eight surgical phases and a transition phase. These are: \textit{Preparation}, \textit{Puncture}, \textit{Positioning of the Guide Wire}, \textit{Pouch Preparation and Catheter Placement}, \textit{Catheter Positioning}, \textit{Catheter Adjustment}, \textit{Catheter Control}, and \textit{Closing}. Transition phases are defined at phase borders as instant stops, breaks, talks, or behaviors that could belong to both surgical phases and are not considered during the training and testing. These few seconds long relaxation periods are proposed to address ambiguity in phase annotations.

\subsection{Temporal Relation}
\label{sec:exp_temp}
\textbf{Problem:} For successful and robust phase recognition, long-distance temporal information must be aggregated to the estimation step at the current time frame. Even for humans, trying to estimate a surgical phase from a short data window without a prior knowledge would be challenging or even may not be possible in some cases. With our choice of TCN architecture, we aimed to mitigate this problem. However, convolutional layers cannot access previous frames outside of the mini-batch, preventing them from leveraging this crucial information. This factor limits the receptive field of the proposed models to batch size. Previous estimation steps can contain relevant information for the current estimation step.
\newline\textbf{Proposed Solution:} To address this problem, we designed an experiment with three settings: (1) Two-Stage TCN. In this setting, the output of the audio and visual branches are concatenated and used as input to the linear layer. The Two-Stage TCN network uses this input for surgical phase recognition; (2) Positional Encodings~\cite{vaswani2017attention}. In addition to the previous setting, we added positional encodings to the output of the linear layer via summation. We aimed to provide timely information to the current estimation, thus, distinguishing similar-looking inputs via their time order. For example, the beginning and the end of each operation have similar audio and visual characteristics and belong to different classes, e.g. \textit{Preparation} and \textit{Closing}. We claim that it would be easier to differentiate these classes by knowing their position within the ongoing operation; (3) Auto-regressive delayed estimations. In this setting, we used the phase estimations from the previous mini-batch to create an additional memory-like feature instead of positional encodings. We added this vector to the output of the linear layer via summation, see Figure~\ref{fig:model}. As a result, we attempted to keep track of both position and phase order simultaneously. We claim that using the previous estimation includes necessary time order information as positional encodings and additional phase order information. Similar to the previous analogy, it would be easier to classify an input at the end of the operation as \textit{Closing} by knowing that the previously estimated phase was the \textit{Catheter Control}.   
\newline\textbf{Implementation Details:} After our initial experiments in the first setting with single, two, and three-stage TCN models, we observed the best results with the two-stage model and used this model for all experiments. The two-stage model has 2.8 million parameters. We performed our experiments using class-weighted cross-entropy loss and Adam~\cite{kingma2014adam} optimizer with weight decay $1E-6$, learning rate $9E-6$, and batch size $512$. Our method was implemented in PyTorch and our models were trained on a single NVIDIA RTX 3090 Ti 24 GB GPU. 
\newline\textbf{Evaluation \& Discussion:} In the evaluation, we used frame-wise accuracy and weighted F1-score. Our results are presented in Table~\ref{tab:temporal}. We observed lowest accuracy in the first setting with the vanilla two-stage TCN network. In parallel to our hypothesis, we achieved significant performance improvement with the addition of temporal connections. By using positional encodings, we reached approximately $8.5$ points higher accuracy. By using auto-regressive delayed connection, we achieved $3.5$ points further increase in the accuracy. In F1-score results, similar results can be seen. 
\begin{table}[h]
  \caption{Phase recognition results of three settings of the temporal model: (1) Two-Stage TCN, (2) Addition of Positional Encodings (3) Addition of past delayed estimation.}
  \label{tab:temporal}
  \centering
  \begin{tabular}{l c c}
    \toprule
    \textbf{Temporal Connection} & \textbf{Accuracy} & \textbf{F1-Score} \\
    \midrule
    Two-Stage TCN             & $67.94 \pm 8.67$ & $68.62 \pm 9.22$        \\
    + Positional Encoding    & $76.48 \pm 4.92$ & $76.15 \pm 5.91$          \\
    + Delayed Estimation                & $79.99 \pm 7.57$ & $80.74 \pm 6.47$        \\
    \bottomrule
    \end{tabular}
\end{table}

\subsection{Class-Imbalance}
\label{sec:class-imbalance}
\begin{figure}[!b]
  \centering
  \includegraphics[width=\linewidth]{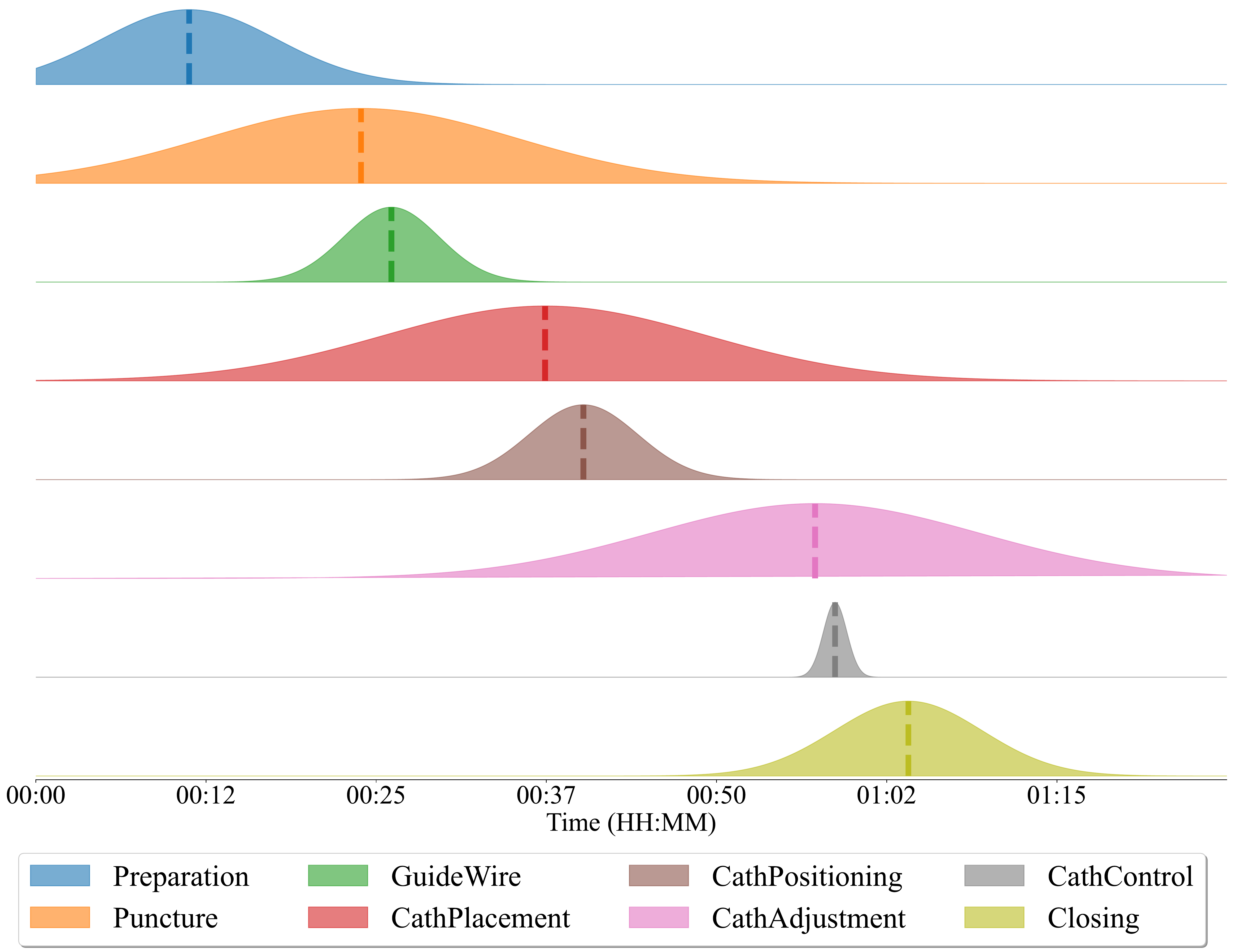}
  \caption{Cumulative mean (dashed line) and Gaussian density estimation (shaded area) of the durations for eight surgical phases in the PoCaP Corpus.}
  \label{fig:phase_dist}
\end{figure}
\begin{figure*}[t]
  \centering
  \includegraphics[width=\textwidth]{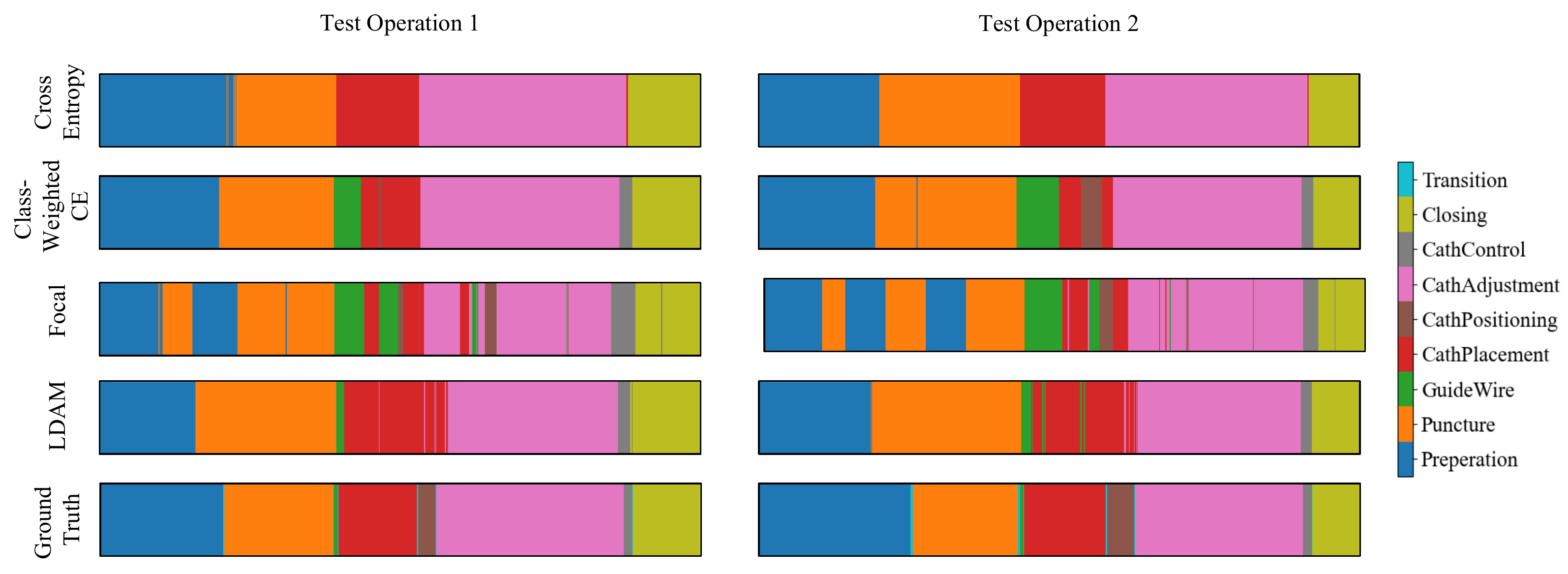}
  \caption{Visualization of estimated surgical phases and ground truth labels of two port-catheter placement operations.}
  \label{fig:ribbon}
\end{figure*}
\textbf{Problem:} Although it is equally important from clinical perspective to recognize all phases robustly, the durations of surgical phases of the port-catheter placement procedure and the corresponding data are not evenly distributed. This class-imbalance problem possesses a challenge during both training and testing. Figure~\ref{fig:phase_dist} shows distribution of durations via Gaussian density estimation for each surgical phase and cumulative means in the PoCaP Corpus. The phases \textit{Guide Wire} (green), \textit{Catheter Positioning} (brown), and \textit{Catheter Control} (grey) have considerably shorter durations when compared to \textit{Puncture} (orange), \textit{Catheter Placement} (red), and \textit{Catheter Adjustment} (pink) phases. Thus, it is difficult to recognize these phases.
\newline\textbf{Proposed Solution:} Several strategies are proposed to mitigate class-imbalance problem in various research topics, including re-weighting of classes and modified loss functions. In this experiment, we considered these techniques with our framework and tested following settings: (1) Cross-entropy (CE) loss. Although an intuitive decision given the strong class-imbalance problem would be the class-weighted cross-entropy loss, we used this setting to illustrate the effects of the re-weighting method in the next step; (2) Class-weighted cross-entropy. In this case, class weights are calculated as inverse frequencies; (3) Focal loss~\cite{lin2017focal}. In this setting, we aimed to give less importance to well-classified phases; (4) Label-distribution-aware margin (LDAM) loss~\cite{cao2019learning}. We attempted to stimulate larger margins for short duration phases in this setting.
\newline\textbf{Implementation Details:} In all experiments in this section, we used the two-stage TCN model with the delayed estimation connection. We kept all other variables and hyperparameters except loss functions the same as in the previous section. In the focal loss experiment, we tested different $\gamma$ values and reported the best results with $\gamma = 2$.
\newline\textbf{Evaluation \& Discussion:} In the evaluation, we used frame-wise accuracy, weighted F1-score, and ribbon plots. Our results are presented in Table~\ref{tab:class-imbalance} and Figure~\ref{fig:ribbon}. We achieved the highest frame-wise accuracy with cross-entropy loss, however, as depicted in the first row of the Figure~\ref{fig:ribbon}, this setting could only correctly classify dominant phases \textit{Preparation} (blue), \textit{Puncture} (orange), \textit{Catheter Placement} (red), \textit{Catheter Adjustment} (pink), and \textit{Closing} (yellow). This was an expected behavior since we handled all phases equally without considering the label distribution of the dataset. Although the overall classification rate is high, this is not a clinically desired output. In the second setting, we used class-weighted cross-entropy and could recognize \textit{Guide Wire} (green) and \textit{Catheter Control} (grey) phases additionally. \textit{Catheter Positioning} (brown) is either mis-classified or entirely missed and \textit{Guide Wire} (green) is over-emphasized in all test data. The second row of the Figure~\ref{fig:ribbon} shows the results for this setting. In the third experiment, we observed similarly over-emphasized short phases \textit{Guide Wire} (green) and \textit{Catheter Control} (grey) phases and many phase shifts with the focal loss function. In addition, we obtained the lowest accuracy and F1-score results in this experiment. This could be resulted from probabilities of class estimations being not significantly different for easy and hard samples. Example results are illustrated in the third row of Figure~\ref{fig:ribbon}. In the final setting, we achieved the most consistent results with LDAM loss. In this case, we could recognize all phases except \textit{Catheter Positioning} (brown) consistently and achieved $2.5$ points better accuracy than class-weighted cross-entropy loss. Results are shown in the fourth row of the Figure~\ref{fig:ribbon}. We think that this setting provides the most robust results for possible clinical applications. \textit{Catheter Positioning} (brown) phase is consistently misclassified in all experiments. We claim that this is caused by this phase being both very difficult to distinguish from neighboring phases and having a short duration. In contrast to this phase, another short phase \textit{Catheter Control} (grey) phase has a very distinctive X-Ray setting, which makes it easier to recognize, thus, it is recognized better in all experiments.

In future studies, we would like to focus on \textit{Catheter Positioning} (brown) phase specifically. Moreover, we would like to experiment with the individual effects of each microphone channel and X-Ray input. Physicians and assistants have different tasks during an intervention and their contributions would be different. An ambient microphone channel typically captures all background sounds in the OR and has a noisy input. The X-ray channel provides very sparse but informative data. Thus, understanding the contribution of multi-modal data is a promising research direction. Finally, we would like to test our approach with different interventions, medical institutes and languages.

\begin{table}
  \caption{Phase recognition results of four settings with different loss functions proposed for class-imbalance problem.}
  \label{tab:class-imbalance}
  \centering
  \begin{tabular}{l c c}
    \toprule
    \textbf{Loss Function} & \textbf{Accuracy} & \textbf{F1-Score} \\
    \midrule
    Cross Entropy             & $84.82 \pm 6.76$ & $82.24 \pm 6.58$        \\
    Class-Weighted CE    & $79.99 \pm 7.57$ & $80.74 \pm 6.47$         \\
    Focal                & $70.19 \pm 4.90$ & $58.35 \pm 3.39$        \\
    LDAM             & $82.56 \pm 3.21$ & $81.30 \pm 3.89$        \\
    \bottomrule
  \end{tabular}
\end{table}

\section{Conclusion}
\label{sec:conc}

In this work, we introduce the PoCaPNet, a framework for surgical phase recognition using speech and X-Ray images. Our study is based on audio features extracted from three different microphone channels using the wav2vec 2.0 XLSR-53 model and visual features extracted from X-Ray images using the TorchXRayVision model. To the best of our knowledge, we are the first to employ speech data from the entire intervention and X-Ray data in combination for the surgical phase recognition task. Aggregating long-term temporal information and learning with class-imbalanced data were the two biggest problems in our study. We proposed using delayed estimation in an auto-regressive manner to integrate past temporal information into the current classification step and LDAM loss to address the class-imbalance problem. Our experimental results show significant performance improvement with these additions. This study shows proof-of-concept for using speech data in an SPR task. Our results encourage the development of many new applications such as interactive intelligent assistants in ORs.

\section{Acknowledgements}
We gratefully acknowledge funding for this study by Friedrich-Alexander-University Erlangen-Nuremberg, Medical Valley e.V. and Siemens Healthineers AG within the d.hip framework.
\newpage
\bibliographystyle{IEEEtran}
\bibliography{mybib}

\end{document}